\documentclass[aps,prl,preprint,groupedaddress]{revtex4}

\usepackage{amssymb,amsmath,amsbsy}
\usepackage{mathrsfs}
\usepackage[dvips]{graphics}
\usepackage{epsfig}

\def\be{\begin{eqnarray}}
\def\ee{\end{eqnarray}}
\def\bea{\begin{eqnarray*}}
\def\eea{\end{eqnarray*}}

\def\tilde{\widetilde}

\def\centeron#1#2{{\setbox0=\hbox{#1}\setbox1=\hbox{#2}\ifdim
\wd1>\wd0\kern.5\wd1\kern-.5\wd0\fi
\copy0\kern-.5\wd0\kern-.5\wd1\copy1\ifdim\wd0>\wd1
\kern.5\wd0\kern-.5\wd1\fi}}
\def\ltap{\;\centeron{\raise.35ex\hbox{$<$}}{\lower.65ex\hbox{$\sim$}}\;}
\def\gtap{\;\centeron{\raise.35ex\hbox{$>$}}{\lower.65ex\hbox{$\sim$}}\;}

\newcommand{\newc}{\newcommand}
\newc{\qbar}{{\overline q}}
\newc{\Kahler}{K\"ahler }
\newc{\deltaGS}{\delta_{\rm GS}}

\begin{document}

\preprint{
\vbox{\vspace*{2cm}
      \hbox{December, 2017}
}}
\vspace*{3cm}

\title{Exploring Non-Universal Gauge Mediation}
\author{Linda M. Carpenter}

\affiliation{Department of Physics and  \\
   Center for Cosmology and AstroParticle Physics, CCAPP \\
   Ohio State University 191 W Woodruff Ave \\
   Columbus, OH U.S.A. \\
   \vspace{1cm}}

\begin{abstract}

I explore the phenomenology and models of Gauge Mediation with a mixed spectrum of Dirac and Majorana gauginos. Scalar sfermion masses are generated using a mixture of Supersoft and Minimal Gauge Mediated communication mechanisms. I build simple models where Dirac and Majorana gauginos arise from distinct SUSY breaking sectors. I explore the phenomenology of the gauge mediated parameter space, identifying candidate NLSP's and possible mass hierarchies.  The phenomenological focus is on 'collider safe' models where the gluino is Dirac. Within this model space I find a range of complex SUSY spectra and decay chains.

\end{abstract}

\maketitle

\section{Introduction and Motivation}

Low energy supersymmetry (SUSY) remains a compelling possibility for weak scale Beyond the Standard Model physics.  Among the leading formalisms for the communication of SUSY breaking is Gauge Mediation, which allows predictive, calculable models, and is flavor bind \cite{Dine:1993yw}\cite{Dine:1994vc}\cite{Dine:1995ag}.  To implement Gauge Mediation most minimally, heavy fields in full SU(5) multiplets, the messengers, couple both to the spurions of SUSY breaking and to the MSSM fields through Standard Model gauge interactions. The messengers have a supersymmetric and non-supersymmetric mass, inducing gaugino masses at the one loop level and scalar mass squared's at the two loop level. The resultant MSSM masses are all  proportional to a single mass scale $F/M$, the ratio of non-supersymmetric to supersymmetric messenger mass parameters. 

Since the ratios of masses are proportional to powers of gauge coupling, particles charged under QCD are much heavier than those that are not.  In addition, all masses are proportional to one scale, hence once an exclusion has been made of light particles mass, the masses of all the other particles are constrained to be quite heavy.  Given current experimental exclusions, a typical minimal mass spectrum is a light neutralino, several hundred GeV weakinos and sleptons, multi-TeV gluino mass and very heavy squarks.  One notes that this spectrum is not very different from the pattern of masses in anomaly mediation or much of standard  mSUGRA space.  Though this realization of Gauge Mediation is simple and elegant, the limitation of a single parameter theory prevent it from being likely.

Supersoft Supersymmetry is and alternate method of the communication of SUSY breaking \cite{Fox:2002bu}. In this mechanism, the low energy theory contains extra chiral fields that are adjoints under the SM gauge groups, and a broken U(1) gauge symmetry in the hidden sector.  Due to interactions with the hidden sector $U(1)$, the gauginos receive a Dirac mass that 'marries' them to the chiral adjoints. These masses constitute three distinct parameters. The scalar masses are then generated a loop factor below gauginos masses, and thus the scalar masses are roughly an order of magnitude lighter than gaugino masses. The features of the Supersoft spectra are quite distinct from that of Minimal Gauge Mediation. With small additions to the minimal theory- the gauginos may be made arbitrarily heavier than the scalars, only the sfermions and Higgsino-like weakinos remain light (heavy binos and winos are not collider accessible) \cite{Carpenter:2016lgo}. The phenomenology is considered 'collider safe' as the heavy gluino is not expected to be produced at the LHC, and  due to decreased production cross sections for squarks, lower mass bounds are relaxed \cite{Kribs:2012gx}. In addition, sparticle decay chains may be quite unusual.

Both Minimal Gauge Mediation (MGM) and Supersoft mechanisms may be built into a much larger theoretical construct, that of General Gauge Mediation (GGM).  The  general definition of this mechanism is that Gauge Mediation is defined as any mediation scheme such that when the Standard Model gauge coupling are turned off, the MSSM masses go to zero \cite{Meade:2008wd}.  Supersoft models may be implemented as Gauge Mediated models using a set of mediating messengers with charges under the SM and the hidden U(1), which couple the gauginos to the chiral adjoint fields. Minimal Gauge Mediation may be extended to include multiple types of messengers and multiple SUSY breaking fields or sectors. General Gauge Mediation admits six mass parameters into the particle spectrum, one for each gaugino, and one for each SM gauge group appearing in the scalar masses.  In this formalism, model building is extremely rich, messengers may either participate in SUSY breaking or not, they may carry additional gauge quantum numbers or not, and hidden sectors may be strongly coupled or not.  The phenomenological bounds on GGM implementations has been studied generally, for example in \cite{Rajaraman:2009ga}\cite{Grajek:2013ola}\cite{Carpenter:2008he}. However almost all of the studied models explore the regime of Majorana gauginos, which does not posses the collider the 'super-safeness' of supersoft models.

In this work I propose to generate and study an extremely general GGM particle spectrum containing both Dirac gauginos from Supersoft models, and the Majorana gauginos arising from a General Gauge Mediated spectrum. This gaugino admixture is the most general form of General Gauge Mediation, and the spectra and constraints will be quite distinct from the Anti-split spectrum of Supersoft models, and from General Gauge Mediation with Majorana gauginos alone.  Of particular phenomenological interest are regions where the gluinos are Dirac, but much of the other spectrum arises from General Gauge Mediation.  Though there has been some phenomenological work on adding Dirac gluinos to restrictive minimal models, \cite {Dudas:2013gga} or, for an mSUGRA example, \cite{Busbridge:2014sha},  the full theoretical parameter space has not yet been explored. Non-Universal Gaugino Masses, in the GGM formalism allows maximal spectral complexity while maintaining some of the collider safe features of Supersoft models.

In model space where the gluino is fully Dirac, the squarks retain some portion of their masses from the Supersoft mechanism. It is this region of the multi-parameter space is favored by current collider data.  The current run of LHC has places excludes gluinos decaying in standard jet plus missing energy channels is ~2TeV \cite{ATLAS}, while projections for discoverability in the 14 TeV run top out under 3 TeV. This suggests that a Dirac gluino of a few TeV is perfectly consistent with LHC data.  Current exclusions for squarks are in the TeV range but may be substantially lower for models with Dirac gluinos.  Mass spectra with relatively low lying Dirac gluinos predict lightish squarks, and hence a possible compressed spectrum.  The flexibility of the GGM spectrum for the remaining sparticles allows for a range of NLSP candidates and hence highly non-standard particle decay chains.

This paper proceeds as follows, Section 2 reviews the formalism of minimal General Gauge Mediation and Supersoft models.  Section 3 demonstrates the formalism for simple weakly coupled SUSY breaking models which could generate the Non-Universal gaugino spectrum.  Section 4 explores mass spectra and collider phenomenology for a range of low energy parameters,  Section 5 concludes.

\section{Review of Low Energy Operators}
The simplest implementation of Gauge Mediation requires the introduction of sets of messenger fields $M_j$ which are charged under the three SM gauge groups. The field content is chosen to be non-anomalous, often simple messenger sectors consist of sets of vector-like messengers $M_j$ and $\overline{M_j}$.  These messengers couple to MSSM scalars and gauginos with normal gauge couplings, and they also couple to sources of super-symmetry breaking.  The simplest superpotentials couple the messengers to one or more SUSY breaking spurions, $S_i$ where S's have vevs and SUSY breaking F terms, $S_i= v_i + \theta^2 F_i$.  This coupling gives the messengers both  holomorphic and non-holomorphic masses. The superpotential is

\be
W = \lambda_{ij} S_i (M\overline{M})_j
\ee

yielding mass terms

\be
L =  \lambda_{ij} v_i (\Psi_M \overline{\Psi_{\overline{M}}})_j + \lambda_{ij} F_i (\tilde{m}\tilde{\overline{m}})_j
\ee

The most minimal model uses a single spurion field and one set of messengers in a fundamental representation of SU(5).  The entire SUSY mass spectrum depends on a single parameter, $\Lambda = F/v$. For example, the gaugino masses arise at one loop
\be
m_{i}=\frac{\alpha_i}{4\pi}\Lambda
\ee

Most generally gauge mediation has six parameters, three determine gaugino masses while, three more determine the SU(3), SU(2), and U(1) contributions to the scalar masses. This may be achieved by coupling various SUSY breaking fields couplings to multiple sets of messengers.  The phenomenological  spectrum may vary quite considerably.  Very generally, the gluino, wino and bino masses are determined by three parameters

\be
m_{g}=\frac{\alpha_3}{4\pi}\Lambda_G;\quad m_{w}=\frac{\alpha_2}{4\pi}\Lambda_w;\quad m_{b}=\frac{\alpha_1}{4\pi}\Lambda_b~;
\ee

while scalar masses arise at two loops, and depend on three more parameters

\begin{equation}
m_s^2= 2\left(C_3 (\frac{{\alpha_3}}{4\pi})^2 \Lambda_3^2 +C_2 (\frac{\alpha_2}{4\pi})^2 \Lambda_2^2 +\frac{Y}{2}^2k(\frac{\alpha_1}{4\pi})^2\Lambda_1^2~ \right).
\end{equation}

Where $C_3$ and $C_2$ are 3/4 and 4/3, the quadratic Casimir's of the SU(3) and  SU(2) gauge groups, and k is 5/3.  As the gaugino masses arise at one loop, and scalar mass squared's arise two loops, we expect that for  $\Lambda$'s of the same order, the scalar and gaugino masses are roughly equivalent.

A quite different way to generate the mass spectrum for MSSM fields is through supersoft SUSY breaking.  In this model we will require the existence of
additional chiral fields $A_i$ which are in adjoints representations of the three SM gauge groups. It will also require a hidden sector U(1) gauge symmetry which gets a SUSY breaking D term vev. The superpotential then contains terms

\be
W = c_i\frac{W^{'}W_i A^i}{\Lambda}
\ee

where $W_i$ are the SM gauge field strengths, $W^{'}$ is the hidden sector U(1) gauge field and  $A_i$ are the chiral adjoints.  The index $i$ runs over the three SM gauge groups.  Gauge indices are contracted between the field strength tensor and the adjoint, while Lorentz indices are contracted between the two field strength tensors. Inserting the D-term vev for $W^{'}$,  Dirac gaugino masses are generated,
\be
c_i \frac{D}{\Lambda}\lambda_i \psi_{Ai}
 \ee
with the MSSM gauginos acquiring masses  $c_i{D}/{\Lambda}$.  There are various options for generating this hidden sector U(1) term as will be discussed later in this work.
Scalar mass-squared's are generated after the gauginos have gotten masses, the scalar masses are finite and one loop level lower than the gaugino masses, with the gauginos or real scalar adjoint running in the loop.  The scalar masses from the Supersoft process are

\be
m_s^2= \frac{C_i \alpha_i {m_{\lambda i}}^2}{\pi} \log (\frac{\delta_i }{ m_{\lambda i}})^2
\ee

where the gaugino masses are $m_{\lambda i}$, $\delta_i$ is the mass squared of the real part of the adjoint field, and $C_i$ are the Casimirs of the fields.  In simple supersoft models, the gauginos are roughly an order of magnitude heavier than scalars, with a scalar to gaugino masses ratio of

\be
\frac{m_s}{m_\lambda}= \sqrt{\frac{2 C_i\alpha_i }{\pi}\log (\frac{\delta_i }{ m_{\lambda i}})}
\ee

It is normally expected that  the real part of the scalar adjoint is twice the gaugino mass. However, various SUSY breaking operators arise to change the masses of the real and imaginary parts of the adjoint field. Thus the exact value of the real adjoint mass will be model dependent \cite{Csaki:2013fla}   \cite{Carpenter:2010as} \cite{Carpenter:2015mna} and \cite{Nelson:2015cea}. There is a special effect as one varies the real scalar mass, if $\delta_i$ nears $m_{\lambda i}$, the gaugino masses make no contribution Supersoft scalar masses and the gauginos become arbitrarily heavier than the scalars.

The Supersoft operators may arise from a form of General Gauge Mediation.  The Supersoft formalism may be embedded into 'semi-direct' gauge mediation, where a set of messengers is charged both under the SM gauge groups and a hidden sector gauge group without participating directly in supersymmetry breaking \cite{Seiberg:2008qj}.  One can couple the new SM  adjoint fields $A_i$ directly to sets of messenger fields, here called T,  which are charged both under the Standard Model gauge groups and the hidden sector U(1) gauge symmetry. A simple superpotential is

\be
W_T= m_T T\overline{T}+ y_i\overline{T}AT
\ee
Here we have given the messengers a supersymmetric mass term.  The messengers also have a non-holomorphic mass term resulting from the SUSY breaking D-term of the hidden sector U(1) field,  generating a one-loop masses for gauginos,

\be
m_{\lambda_i} = \frac{g_i}{16\pi^2}\frac{y_i D}{m_T}
\ee
with the scalar masses arriving at the two loop level.  The spectrum of these models is quite different than Minimal Gauge Mediated models. The 'anti-split' hierarchy of the gauginos and scalars has large implications for phenomenology.

\section{Weakly Coupled Models}

A mixed gaugino spectrum may be implemented using a sets of weakly  couples operators. We begin by building a very simple hidden sector, where the Dirac gauginos arise in a different SUSY breaking sector from the Gauge Mediated sector that yields Majorana gauginos. We will thus require one simple O'Raifeartaigh model with one or more SUSY breaking spurions, and one sector with a gauged U(1) symmetry which must have a non-zero D term at the SUSY breaking minimum.  We begin with the hidden sector, \cite{Carpenter:2010as},  that will generate the Supersoft operators with a simple superpotential

\be
W_1= \lambda X (\phi_{+}\phi_{-}- \mu^2) + m_1\phi_{+}Z_{-} + m_2 \phi_{-}Z_{+}
\ee

Here there exists a gauged U(1) symmetry. The field  X is a charge neutral  SUSY breaking spurion, and the indices on the fields $\phi$ and Z indicate charge under the U(1) gauge group. In this model, the fields $\phi$ get vevs

\be
\phi_{+}^2=\frac{m_2}{m_1}\phi_{-}^2
\ee
\be
\phi_{-}= \sqrt{\frac{m_1}{m_2}\mu^2-m_1^2} \nonumber
\ee
spontaneously breaking the U(1), while the field X gets an F term.

The U(1) D-term is nonzero as long and $m_1$ is unequal to $m_2$, and is proportional to
\be
D = g^{'}(\frac{m_1}{m_2}\mu^2-m_1^2)(\frac{m_2}{m_1}-1)
\ee

The D-term must now be coupled to a messenger sector to give Dirac mass to certain gauginos.  This will require the addition of  messengers which
are charged both under the hidden sector U(1) and some Standard Model gauge groups.  The additional messengers, $T_i$, will require their own supersymmetric mass-term and will couple to the adjoint fields $A_k$.  The simplest model contains messengers which are fundamentals and anti-fundamentals under one or more SM gauge groups.  We thus write the messenger sector
\be
W_T= m_{Tij} T_i\overline{T_j}+ y_{ijk}\overline{T_i}A_k T_j
\ee

If the messengers do not couple to the field X, they will not get a tree level B term, and resulting gaugino masses will dominantly  come from the Supersoft terms.  This may be arranged by invoking and R symmetry.  If the fields X has R charge 2, the $\phi$'s opposite R-charges, the Z's 2 minus the charge of $\phi$, and the messengers R charge 1, the superpotential may be protected.  Notice that we need more than one messenger pair to couple to a single adjoint field for a sensible model.  This is because a breaking of messenger parity is needed generate operators which prevent negative mass squared's for the scalar adjoint fields  \cite{Csaki:2013fla}.

We will now include a SUSY breaking sector which leads to Majorana Masses for some of the gauginos. The simplest thing to do, though aesthetically byzantine, is to couple a different set of messengers to a new SUSY breaking spurion that gets an F term and a vev. To arrange the most general spectrum, we would like to allow messengers charges under different SM gauge groups to have separate couplings to the spurion. One of the simplest sectors to couple to is that of Extraordinary Gauge Mediation \cite{Cheung:2007es}.

For example we may write a superpotential
\be
W_2=YF+ \lambda_{ij}Y\phi_i\overline{\phi_j}+m_{ij}\phi_i\overline{\phi_j}
\ee
Where Y is a SUSY breaking spurion, and $\phi_i$ and $\overline{\phi_i}$ are sets of messengers in simple fundamental representations of the SM gauge groups only.  Unlike the previous sector, this sector requires a messenger parity in order to avoid dangerous one loop level diagrams that could contribute large negative mass squareds to some sfermions through hypercharge D terms.  Instituting this messenger party for various sets of messengers charges under SM gauge groups was explored in \cite{Carpenter:2008wi}.  Sequestering between the two SUSY breaking sectors ensures that R-symmetry breaking this sector will not be easily communicated to the other sector.

A gravitino mass will be generated with value $m_{\tilde{G}} = F/\sqrt{3}M_{P}$.  Where F is the dominated by the highest SUSY breaking F-term in the theory.  The gravitino is always expected to be the Lightest Supersymmetric Particle (LSP) and couples to all sparticles in the theory, however, for $\sqrt{F}$  above $10^6$ GeV the width decay of sparticles into the LSP is highly suppressed and sparticles produced at LHC cannot decay inside of the detector.  This has large consequences for the collider phenomenology of Gauge Mediated models.  For SUSY breaking sectors contained in the superpotential $W_2$, such low F terms are not problematic.  However, technicalities generically arise  when generating a SUSY breaking D-term. Though the D-term is sometimes comparable to the F terms in size, it is often the case that the D-term generated is parametrically smaller than the F-terms arising from many DSB models. In this case, much larger F-terms are needed to generate a D-term large enough to produce gaugino masses in the TeV range, and the decays to the gravitino are suppressed. It is important to keep this in mind when choosing a SUSY breaking sector. In the model above, the D term is proportional to $\vert\phi_+\vert^2-\vert\phi_{-}\vert^2$ with F term of order $\phi_i^2$. It is expected there should be no great cancelation among the vev's, so that the D term remains of respectable size compared to the F terms, thus this model provides the option for moderate F terms and light gravitinos.

\section{Spectra}
As demonstrated above, different sets of messengers may interact differently with the dynamical SUSY breaking sector. We may then produce a spectrum with some Dirac and some Majorana type gauginos.  The phenomenology resultant from the addition Dirac mass contributions to gauginos has been studies in some scenarios, adding a single extra parameter the MSSM spectra (modulo parameters specific to the Higgs sector).  Some example the addition of a Dirac Bino contributions to an Anomaly Mediated spectrum \cite{Carpenter:2005tz}, or the addition of Dirac gaugino contributions to the mSUGRA parameters \cite{Busbridge:2014sha}.  However, General Gauge Mediation provides an enormous freedom of parameters, opening many possibilities for phenomenological spectra.  It also allows some gauginos to be Majorana while others are nearly fully Dirac. The most collider-safe spectra will contain Dirac gluinos, as this provides a suppression of the squark pair production processes, and eliminates squark-gluino production. To demonstrate the richness of the Non-Universal gaugino scenario, I will focus on a sub-set of the Dirac gluino space,  where the wino receives a Dirac mass and the bino mass is generated by a minimal Gauge Mediated process.  There are multiple candidates for NLSP over this space, the spectrum is compressed, and the possible decay chains are quite complex, as I describe below.

\subsection{Majorana Bino Model}
We note something interesting, in order to give masses to the entire MSSM spectrum of scalars through the normal gauge mediated mechanism, one only requires a set of MGM messengers which have SM hypercharge.  The only massless fields are then the wino and gluino, and this masslessness may be cured by invoking the Dirac gaugino mechanism for these fields.  The gaugino mass parameters are then
\be
m_b= \frac{\alpha_1}{16\pi^2}\Lambda_b; m_w= \gamma_2\frac{D}{M};  m_g= m_D = \gamma_3\frac{D}{M}
\ee
with scalar masses
\be
m_{lr}^2=2 \frac{5}{3}{\frac{Y}{2}}^2\frac{\alpha_1^2}{(16\pi^2)^2}\Lambda_1
\ee
\be
 m_{qr}^2=2 \frac{5}{3}{\frac{Y}{2}}^2\frac{\alpha_1^2}{(16\pi^2)^2}\Lambda_1 + \frac{C_3\alpha_3 m_g^2}{\pi}Log(\frac{\delta_3^2}{m_g^2}) \nonumber
\ee
\be
 m_{ll}^2=2 \frac{5}{3}{\frac{Y}{2}}^2 \frac{\alpha_1^2}{(16\pi^2)^2}\Lambda_1+ \frac{C_2\alpha_2 m_w^2}{\pi}Log(\frac{\delta_2^2}{m_w^2}) \nonumber
\ee
\be
m_{ql}^2=2 \frac{5}{3}{\frac{Y}{2}}^2\frac{\alpha_1^2}{(16\pi^2)^2}\Lambda_1 + \frac{C_2\alpha_2 m_w^2}{\pi}Log(\frac{\delta_2^2}{m_w^2}) +  \frac{C_3\alpha_3 m_g^2}{\pi}Log(\frac{\delta_3^2}{m_g^2}) \nonumber
\ee

where $C_i$'s are the group Casimir coefficients and $\delta_i$ are the non-supersymmetric masses of the heavy adjoints.

This simple model has three parameters $(m_g, m_w, \Lambda_1)$. However, the parameter space of mass spectra is quite different from the three parameter models  of Supersoft Mediation, or typical three parameter models of General Gauge Mediation with Majorana gauginos.  We will now study the resultant spectra. 

Notice that for choices of intermediate gluino masses in the few TeV range, but relatively heavy binos, the mass spectrum can be heavily compressed.  As a benchmark, consider the point where the wino mass is fixed to be 500 GeV with a $1.2$ TeV $\mu$ term. To begin, we must determine the particle content of the next to lightest SUSY particle (NLSP), which varies greatly over the parameter space.

\begin{figure}[h]
\begin{center}
\includegraphics[width=3.0in]{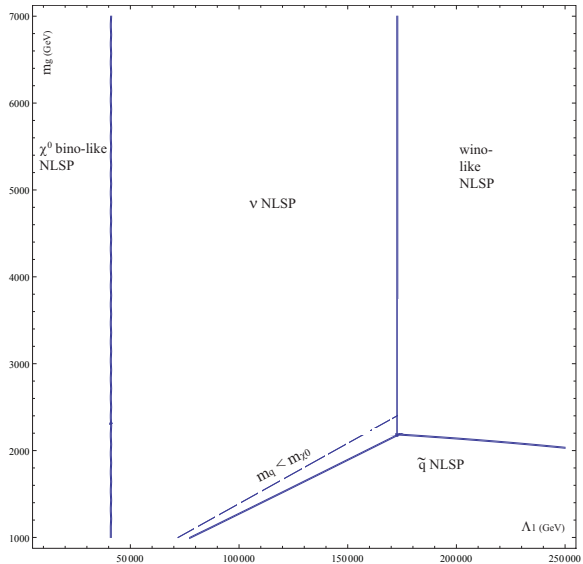}
\end{center}
\caption{NLSP particles in the $m_g, \Lambda_1$ mass plane for $m_w$=500 GeV $\mu$=1.2 TeV }
\end{figure}

The behavior of the superparticle spectrum changes drastically as the U(1) gauge contribution overtakes other contributions to the mass spectrum.  The NLSP content is mapped over the parameter space in Figure 1. We see that for this benchmark, the bino is the NLSP as long as the scale $\Lambda_1$ is low.  For larger values of the bino mass,  the sneutrino becomes the NLSP.  As $\Lambda_1$ increases further, the slepton masses overtake the wino mass parameter, and the NLSP becomes wino-like. There is a peculiar region in parameter space where the squarks become the NLSP.  In this region, the Dirac gluino mass is low, under 3 TeV, so that the squark masses are lighter than the wino mass parameter, while  $\Lambda_1$, is large.  It is unclear if any of this region of parameter space is phenomenologically allowed,  squark masses in this region are contour plotted in Figure 2. There is also a small region of sneutrino NLSP space where the squarks are lighter than the binos, which will complicate the quark decay chains.

\begin{figure}[h]
\begin{center}
\includegraphics[width=3.0in]{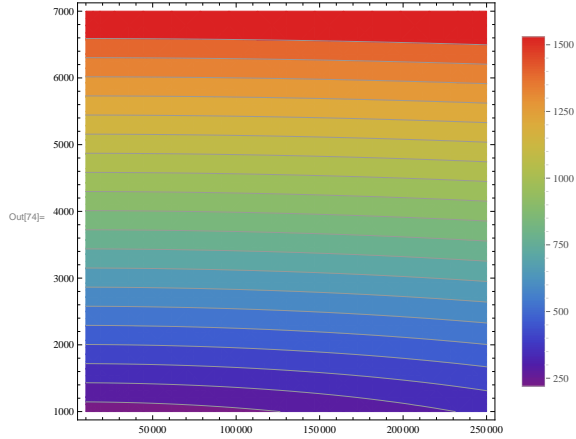}
\end{center}
\caption{Contours of squark masses in the  $m_g, \Lambda_1$ mass plane for $m_w$=500 GeV }
\end{figure}

\begin{table}[h]
\begin{center}
\begin{tabular}{c|c|c|c|c}
sparticle & mass (GeV) bino NLSP  &  sneutrino NLSP & squark NLSP  & wino NLSP\\
\hline
$\tilde{\tau_r}$  & 201 & 576 & 1150 & 1037\\
$\tilde{\nu}$  & 114  & 292 & 578 &  521\\
$\tilde{q_{l}}$  & 868 & 872 & 456 & 778\\
$\tilde{q_{ur}}$  & 866 & 870 & 453& 776 \\
$\mu$  & 1200  & - & -\\
$\chi^{0}_1$  & 110 & 315 & 503 & 500\\
$\tilde {g}$ & 4000 &  4000 & 1900 & 3500\\
\end{tabular}

\caption{Sample spectra from the 4 NLSP candidate scenarios for benchmark values $m_w=500$ GeV, $\mu=1.2$ TeV.}
\label{tab:gauginobounds}
\end{center}
\end{table}

In Table 1, sample spectra from the 4 NLSP benchmark scenarios are shown. Next proceeds a discussion of possible decay chains and collider bounds in the various NLSP scenarios.  In the light gravitino scenario, the NLSP will decay to the gravitino plus a standard model counterpart.  For heavy gravitinos the NLSP will be collider stable. We will now lay out the phenomenology in these to scenarios for each NLSP candidate.

The bino is the NLSP in the region of fairly small $\Lambda_1$ and thus small bino mass of order 100 GeV.  Unstable bino-like NLSP's in a gauge mediated scenario will decay to a photon plus missing energy, $\chi_0\rightarrow \gamma + \tilde{G}$.  Limits in this scenario are unclear, as quotes exclusions for binos decaying in GMSB scenarios require the production of binos along with other MSSM particles, or in decay chains - for example of gluinos (in the Dirac possible collider inaccessible) \cite{Aad:2015hea}. Without significant effort in recasting results a simple limit cannot be drawn. Decays of the sfermions should proceed in to a fermion plus the bino NLSP, $\tilde{q}\rightarrow q + \chi_0 \rightarrow q +\gamma + \tilde{G}$ and  $\tilde{\ell}\rightarrow \ell + \chi_0 \rightarrow q\ell +\gamma + \tilde{G}$.  Limits on the sfermion masses are also unclear in this particular scenario.  Though there are hard photons in the event, pair production of quarks is suppressed in this scenario relative to standard SUSY scenarios.  Searches looking to bound the decaying bino produced in the squark cascade decay chain loose sensitivity due to the squark production cross sections.  Current gluino bounds are the harshest, with gluino searches in the jets plus missing energy channel excluding a maximum of near 2 TeV \cite{ATLAS}.  The gluino decay proceeds through on or off shell squarks down to the bino-like LSP $\tilde{g}\rightarrow q \tilde{q}\rightarrow qq + \chi_0 \rightarrow qq +\gamma + \tilde{G}$.  For inclusive searches looking for gluino decays to jets plus missing energy photons the lower bound is 1.8 TeV \cite{ATLAS2}.

Stable binos will have a much more standard phenomenology, there is no lower mass bound on the stable bino itself.  Sfermion decays will proceed in the standard way, $\tilde{q}\rightarrow q + \chi_0$ and $\tilde{\ell}\rightarrow \ell + \chi_0$.  Here the lower mass bounds on squarks come from jets plus missing energy searches.  Since the squark production cross section is suppressed due to the Dirac gluino's non-participation in the squark production, lower mass bounds on squarks in this regime are expected to be in the range of 800 GeV for very light binos.  Gluino mass bounds will vary but are maximally around 2 TeV from jets plus missing energy searches.

For the case of sneutrino NLSP the stability due to gravitino mass is not an issue. In this region of parameter space the NLSP is heavier relative to the other sparticles then it is in the bino NLSP space.  The neutralino will decay with a pure missing energy signature $ \chi_0 \rightarrow \tilde{\nu}  \nu$, while charginos will decay to a charged fermion and the sneutrino, for example $ \chi^{+} \rightarrow \ell^{+} \tilde{\nu}$.  There are likely no strong bounds on the charginos mass in this case as the spectrum of gauginos and sfermions is quite compressed in this region. The sleptons will decay to a lepton and gaugino, then proceed down the decay chain,  $\tilde{\ell}\rightarrow \ell  \chi_0 \rightarrow \ell  \tilde{\nu}  \nu $,  or, $\tilde{\ell}\rightarrow \nu + \chi^{+/-} \rightarrow \nu \ell^{+/-} \tilde{\nu}$.  The squarks have a quite non-standard decay chain, for example,  $\tilde{q}\rightarrow q  \chi_0 \rightarrow q  \tilde{\nu}  \nu $ or $\tilde{q}\rightarrow q  \chi^{+/-} \rightarrow q  \tilde{\nu}  \ell^{+/-}$.  One expects no strong bounds here due to both the suppression of the squark production and the compression of the spectrum.  The gluinos decay $\tilde{g}\rightarrow q \tilde{q}\rightarrow qq \chi$. With the charged or neutral gaugino decaying down to the sneutrinos.  Here there may be some softening of the bound as the compressed spectrum ensures that some quarks in the event are soft but we may expect gluino mass bounds between 1 and 2 TeV.

Wino-like NLSP is the next scenario to study.  Here the wino-like content ensures a mass degeneracy between the lightest neutralino and chargino.  In the case of light gravitinos, we expect to see the NLSP decay to decay inside the detector.  Bounds on the decaying wino-like NLSP exist between 115 and 370 GeV \cite{Aad:2015hea}.  However, this is below the wino mass thresholds for our benchmark points. The decays proceed in manner similar to the bino-like NLSP, bounds on decaying sfermions in this scenario are not harsh especially given the compressed spectra.   Bounds on the gluino are expected to reach to be 1.8 TeV as before.
Collider stable winos are quite difficult to detect.  Collider-stable winos are very difficult to detect. Due to the mass compressions of the charged and neutral states, the chargino decay products are soft.  Thus both the lightest chargino and neutralino appear to the detector as missing energy.  Direct searches for this scenario look for missing energy produced in conjunction with a recoiling gauge boson emitted as initial or final state radiation \cite{Han:2013usa}\cite{Anandakrishnan:2014exa}. These searches, however, are loose sensitivity to the high wino masses of this benchmark point.  The decay chains of the sfermions and gluino proceed as in the case of a stable wino.  However, given the compression of the spectrum in this region, lower bounds on squarks likely do not exist. It is unclear if the maximum gluino bounds are softened.

Finally the squark NLSP region may be addressed. In the light gravitino scenario unstable squarks will decay to a gravitino and squark. The collider signature for quark pair production is thus two jets plus missing energy, the same signature as for a typical MSSM squark decaying to a massless gluino.  The suppression of squark pair production in this scenario might give one hope that some of the squark NLSP region remains viable.  However, in this benchmark scenario, the mass of the lightest squark remains under 500 GeV. It is therefore likely that this region is excluded by current collider searches. It is possible, however that in a similar benchmark scenario with heavier winos, the right handed squarks may be heavy enough to be viable, while still remaining the NLSP.  Collider stable squarks are heavily constrained by searches for long lived charged particles. Searches for R hadrons demand long-lived up type squarks must be heavier than 1 TeV with normal MSSM production cross sections. It is unlikely that accounting for the drop in squark production cross section, the constraints can accommodate squark masses as low as 500 GeV. The squark NLSP scenario seems disfavored in this benchmark point but could perhaps be made marginally acceptable in a slightly different benchmark scenario for the Non-Universal gaugino mass scenario.

\section{Conclusions}
I have presented the formalism for General Gauge Mediated Models that contain a Non-Universal spectrum of gauginos, some of which are Dirac and some of which are Majorana particles.  I have demonstrated that these models may be built simply by exploiting multiple SUSY breaking sectors with distinct sets of messengers.  Using a set of low energy operators I have laid out the general spectrum of these models which is distinct both from Supersoft models with Dirac gauginos, and from General Gauge Mediated models with Majorana gauginos.

I have focused on the phenomenology of spectra where the gluino has a Dirac mass, as these models exhibit 'collider super-safeness' which GGM models with Majorana gluinos do not possess.  In particular, these models allow lighter squark masses that still avoid LHC bounds.  I have studied a particular  limit of the parameter space, where the U(1) gauge group participates in Minimal Gauge Mediation with a Majorana bino, while the gluino and wino are Dirac.  This generates a peculiar variants for the supersymmetric spectra with a variety of NLSP candidates,  mass compression, and possible complexity in the squark decay chains.

These is much more work to be done in this corner of General Gauge Mediation. On the model building front, it may be possible to build much more concise models of SUSY breaking.  For example, a Dynamical Symmetry Breaking model could be invoked which contains a spontaneously broken U(1) gauge symmetry and a large gaugeable flavor group in which  the SM gauge groups may be embedded. The normal Gauge Mediated messengers might be embedded in the SUSY breaking sector, engaging in  direct mediation, while the U(1) D-term could be coupled to other messengers which generate the gaugino masses.  Such a model would more naturally separate the two messenger sectors into two classes, one which participates in SUSY breaking, and one that does not, while there would be a single SUSY breaking sector.

In addition, there are many phenomenological studies that may be done on these spectra. First there are more benchmark points to study with various combinations of Dirac and Majorana gauginos. In addition 'super-safeness' means many models are un-probed by colliders.  The suppression of the squark pair production cross-section in models with Dirac gluinos means that mass bounds on squarks are quite relaxed compared to models exhibiting heavy Majorana gluinos \cite{Kribs:2013oda}.  In addition, the recasting of typical jets put missing energy squark searches to apply to squarks in Dirac gluino models does not take into account the extreme variations in squark decay chains that could arise in General Gauge Mediation.  Taking these things  together, there may yet be much unbounded parameter space, and the sparticle discovery scenarios for the case of Non-Universal Gauginos may contain extremely non-standard event topologies.

\subsection{Acknowledgements}
The author thanks Yuri Shirman and Yael Shadmi for long ago discussions. This work was supported in part by the US department of Energy grant DE-
SC0011726.


\end{document}